\documentclass[10pt, conference]{IEEEtran}
\IEEEoverridecommandlockouts
\usepackage{cite}
\usepackage{amsmath,amssymb,amsfonts}
\usepackage{algorithmic}
\usepackage{graphicx}
\usepackage{textcomp}
\usepackage{xcolor}
\def\BibTeX{{\rm B\kern-.05em{\sc i\kern-.025em b}\kern-.08em
		T\kern-.1667em\lower.7ex\hbox{E}\kern-.125emX}}

\makeatletter
\newcommand*{\rom}[1]{\expandafter\@slowromancap\romannumeral #1@}
\makeatother
\ifCLASSINFOpdf
\else
\fi
\begin{document}
%
\title{Coordination of PV Smart Inverters Using Deep Reinforcement Learning for Grid Voltage Regulation
}


\author{\IEEEauthorblockN{Changfu Li}
\IEEEauthorblockA{Center for Energy Research\\
University of California San Diego\\
La Jolla, CA USA\\
Email:chl447@ucsd.edu}
\and
\IEEEauthorblockN{Chenrui Jin, Ratnesh Sharma}
\IEEEauthorblockA{Energy Management\\
NEC Laboratories America\\
San Jose, CA USA\\
Email:$\{$cjin, ratnesh$\}$@nec-labs.com}
}


%


\maketitle

\begin{abstract}
Increasing adoption of solar photovoltaic (PV) presents new challenges to modern power grid due to its variable and intermittent nature. Fluctuating outputs from PV generation can cause the grid violating voltage operation limits. PV smart inverters (SIs) provide a fast-response method to regulate voltage by modulating real and/or reactive power at the connection point. Yet existing local autonomous control scheme of SIs is based on local information without coordination, which can lead to suboptimal performance. In this paper, a deep reinforcement learning (DRL) based algorithm is developed and implemented for coordinating multiple SIs. The reward scheme of the DRL is carefully designed to ensure voltage operation limits of the grid are met with more effective utilization of SI reactive power. The proposed DRL agent for voltage control can learn its policy through interaction with massive offline simulations, and adapts to load and solar variations. The performance of the DRL agent is compared against the local autonomous control on the IEEE 37 node system with thousands of scenarios. The results show a properly trained DRL agent can intelligently coordinate different SIs for maintaining grid voltage within allowable ranges, achieving reduction of PV production curtailment, and decreasing system losses.  

\end{abstract}


%
\IEEEpeerreviewmaketitle

\section{Introduction}
The amount of renewable distributed generation (DG) connection has been significantly increasing in recent years due to the technical, economic, and environmental benefits it brings \cite{walling2008}. However, increasing penetration of variable DG like PV generation can cause voltage problems on the distribution power grid \cite{li2018optimal}. 

Conventionally, distribution network operators (DNOs) rely on on-load tap changers (OLTCs) and fixed or switched capacitors to maintain appropriate voltage profile across the network. However, they are limited by number and speed of operations, and insufficient to adapt to highly variable PV production to provide desired voltage regulation. Under latest IEEE 1547 standard \cite{ieee1547_2}, PV generation with a smart inverter (SI) is allowed to participate in grid voltage regulation via various smart functionalities. Those smart functionalities include curtailing PV real power generation (Volt-Watt), injecting or absorbing reactive power (Volt-Var). The implementation has began in California and Hawaii \cite{ca21,rule14h}. Comparing to legacy voltage regulation devices, SIs  provide faster response to grid condition changes for voltage regulation. However, the commonly used Volt-Var function defined in \cite{ieee1547_2,ca21,rule14h} is based on local droop curve, with which the SI absorbs/injects corresponding amount of reactive power per local bus voltage. This could result in suboptimal system performance since there is no coordination between different SIs.

To address this concern, many optimization based methods have been proposed to determine optimal dispatch of SIs \cite{emiliano14,guggilam2016,li2019}. Reference \cite{emiliano14} leverages semi-definite programming relaxation for efficiently solve an optimization of SI real and reactive power. An Alternating Direction Method of Multipliers based algorithm is developed in \cite{guggilam2016} for optimal SI reactive power dispatch and voltage regulation. Reference \cite{li2019} studies coordination of OLTCs and SIs. A linearization technique is proposed to relate bus voltage with controllable variables including OLTC tap position and SI reactive power. However, large computation time to solve those optimizations limits the ability of SIs to respond to fast disturbances caused by moving clouds. 

The success of reinforcement learning (RL), especially deep reinforcement learning (DRL) in various fields including AlphaGo \cite{alphago17}, ATARI games\cite{arcade13}  and robotics \cite{robotics15}, has attracted interest of power and energy community. There have been several works on applying RL/DRL for intelligent control and operation in power grid. Deep Q network (DQN) is used in \cite{diao19} for controlling voltage setting points of generators to maintain acceptable system voltage in response to load variations and line outage. Reference \cite{yang19} dispatches SIs, OLTCs, and capacitors at different time scale for distribution grid voltage control. Optimization is used for fast-timescale dispatch of SIs while slow-timescale OLTCs and capacitors are handled by DQN. A distributed Q-learning is implemented to coordinate generators, OLTCs, and capacitors in \cite{xu12} for optimal reactive power dispatch. Batch reinforcement learning is applied to achieve cooperation of OLTCs for voltage regulation \cite{xu18}. Coordination between OLTCs and capacitors are studied with policy gradient method for voltage violation mitigation and operation cost reduction \cite{wei2019}. Reference \cite{john2004} coordinates OLTCs, capacitors and generators to meet operation limits with Q-learning.   

In all the works discussed above \cite{diao19,yang19,xu12,xu18,wei2019,john2004}, RL/DRL is used for control of devices with discrete settings (generator voltage setting point, OLTC tap position, capacitor tap position). As explained before, SIs are more suitable for accommodating frequent PV generation fluctuations due to their fast response speed in comparison with legacy voltage regulation devices; however, the outputs of the SI are continuous. Discretization of SI outputs could result in more PV generation curtailment due to SI capacity limit. Morever, discretization of SI outputs will face the curse of dimensionality. Performing even a coarse discretization of the SI reactive power output ([-1,1]) of 9 steps with only 5 SIs will result in a $9^5=59049$ action space. Therefore, the deep deterministic policy gradient (DDPG) with actor-critic is adapted from \cite{timothy2016} to handle the continuous control of SIs.

To the best knowledge of the authors, this is the first work applying DRL to coordinate SIs with continuous outputs. After proper training, the DDPG agent will be able to generate timely control decisions since it only needs one feed-forward step of the trained neural network to produce actions of SIs.
These suggested SI actions will be executed to leverage the fast response speed of SIs to accommodate PV generation fluctuations. The performance of the well-trained DDPG agent is compared against the autonomous Volt-Var function recommended in \cite{ca21} on the IEEE 37 node test feeder. The rest of the paper is organized as follows. Section~\ref{sec:prel} introduces preliminaries of distribution power grid and SIs. Details of the implementation of DDPG for coordination of SIs are presented in Section~\ref{sec:DRL}. Case studies are discussed in Section~\ref{sec:casestudy} following by conclusions in Section~\ref{sec:conc}.

\section{Preliminaries}
\label{sec:prel}
In this section, we review the power flow equations for distribution power systems and how smart inverters can be used for distribution grid voltage regulation.
\subsection{Distribution Power System}
\label{sec:prel1}
A distribution grid with $N+1$ nodes can be represented by a graph $\mathcal{G} :=(\mathcal{N}_0,\mathcal{\xi})$, where $\mathcal{N}_0 := \{0,...,N\} $ is the collection of all nodes, and $\mathcal{\xi}:=\{(m,n)\subset \mathcal{N}_0 \times \mathcal{N}_0\}$ is the collection of edges representing distribution lines of the grid. The distribution grid typically operates radially as a tree and is served by a substation (a.k.a. the root) indexed by $n=0$. The substation can be treated as a slack bus where its voltage magnitude $|V_0|$ and angle $\theta _0$ are tightly regulated as constants. The voltage of the whole grid is governed by the power flow equations,
\begin{align}
\sum_{j=0}^{N}|V_k||V_j|\big(G_{kj}cos(\theta _k - \theta _j)+B_{kj}sin(\theta _k - \theta _j)\big)-P_k=0,
\label{pf_eq1}\\
\sum_{j=0}^{N}|V_k||V_j|\big(G_{kj}sin(\theta _k - \theta _j)-B_{kj}sin(\theta _k - \theta _j)\big)-Q_k=0,
\label{pf_eq2}
\end{align}
where $|V_k|$ and $\theta _k$ are the voltage magnitude and voltage angle at bus $k$, respectively; $G_{kj}$ and $B_{kj}$ are the conductance and susceptance between bus $k$ and $j$, which represent the electrical properties of the line connecting bus $k$ and bus $j$; $P_k$ is the real power injection at bus $k$ and $Q_k$ is the reactive power injection at bus $k$.

\subsection{Smart Inverter for Voltage Regulation}
A PV inverter is a type of electrical device that converts the direct current (DC) output of a solar panel into an alternating current (AC) output, which can be fed into the commercial AC grid through the point of common coupling (PCC). Under the new standards/rules \cite{ieee1547_2,ca21,rule14h}, a PV inverter is required to help grid regulation via defined smart functions; this type of PV inverter is referred as a smart inverter (SI) hereafter. An SI is able to help voltage regulation through modulating real and/or reactive power of the PCC, i.e. it can change the $P_k$ and/or $Q_k$ in (\ref{pf_eq1},\ref{pf_eq2}) if bus $k$  has a PV connection. In this way, the SI can change the voltage for bus $k$ as well as other buses per (\ref{pf_eq1},\ref{pf_eq2}).

\begin{figure}[htbp]
	\centerline{\includegraphics[width=0.45\textwidth]{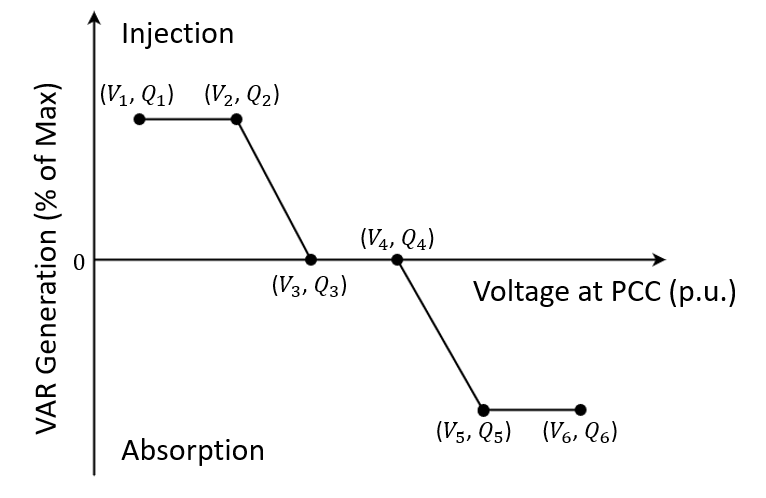}}
	\caption{A typical Volt-Var droop curve of a smart inverter.}
	\label{fig:voltvar}
\end{figure}

\begin{figure}[htbp]
	\centerline{\includegraphics[width=0.45\textwidth]{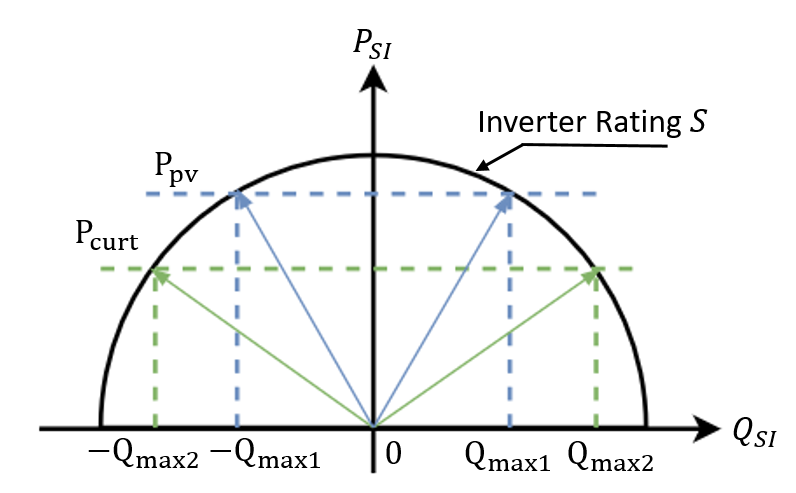}}
	\caption{Smart inverter output curve. The complex power output of the SI is $S_{SI}=P_{SI}+iQ_{SI}$. $S_{SI}$ is constrained by inverter rating $S$, meaning $P_{SI}^2+Q_{SI}^2\leq S^2$. $P_{pv}$ is the available PV real power production determined by instantaneous solar irradiance, $\pm Q_{max1}$ is the corresponding maximum reactive power injection or absorption of the SI. If the real power is curtailed to $P_{curt}$, more headroom is made for modulating reactive power  ($\pm Q_{max2}$).}
	\label{fig:SI}
\end{figure}

A commonly used smart function is a Volt-Var droop curve as shown in Fig.~\ref{fig:voltvar}. There are six different points specifying the shape of the curve, according to which the SI will absorb or inject corresponding amount of reactive power (VAR) based on the voltage at PCC. The real power production of PV can be curtailed to make headroom for VAR generation if the SI reaches its capacity limit as shown in Fig.~\ref{fig:SI}. This scheme is called Volt-Var with VAR priority. With Volt-Var droop curve, every SI operates autonomously based on local voltage without coordination with each other. This brings simplicity in terms of implementation but can also lead to undesired system performance. Since not each bus of the power network is equipped with a SI, some buses may suffer from voltage violations even under the autonomous SI dispatch scheme. Meanwhile, some SIs may use excessive reactive power because of not coordinating with others, resulting in unnecessary PV production curtailment.

\section{Coordination of Smart Inverters Using DRL}
\label{sec:DRL}
\subsection{Reinforcement Learning}
RL, especially DRL, has been shown to be capable of learning by interacting with complicated environments and achieving good performance on difficult control tasks like robot manipulation. Therefore, DRL is chosen to perform grid control since the power grid is complex and dynamic by nature.

In RL, an agent learns through interacting with an environment, $E$. At each time step, the agent receives the state of the environment $s_t$, takes an action $a_t$ and receives a scalar reward $r_t$. The agent learns a policy $\pi$, which maps states to a probability distribution over the actions $\pi:\mathcal{S}\rightarrow\mathcal{P}(\mathcal{A})$. This can be modeled as a Markov decision process with a sate space $\mathcal{S}$, action space $\mathcal{A} = {\rm I\!R}^M$, an initial state distribution $p(s_1)$, transition probability $p(s_{t+1}|s_t,a_t)$, and reward function $r(s_t,a_t)$. $M$ is the dimension of the action space.

The agent uses the policy to explore the environment and generate states, rewards and actions tuples, ($s_1,a_1,r_1,....,s_t,a_t,r_t$). The return of a state is calculated as the total discounted future reward from time step $t$ and onwards, $R_t = \sum_{i=t}^{T}\gamma^{(i-t)}r(s_i,a_i)$, where $\gamma \in [0,1]$ is the discount factor quantifying the importance attached to future rewards. The goal of the agent is to learn a policy that results in maximization of cumulative discounted reward from the start distribution $J = \mathbb{E}_{r_i,s_i\sim E,a_i\sim \pi}[R_1]$.

The action value function is defined as the expected total discounted reward after taking an action $a_t$ in state $s_t$ and thereafter following policy $\pi$:
\begin{align}
Q^\pi(s_t,a_t) = \mathbb{E}_{r_{i\geq t},s_{i\geq t}\sim E,a_{i\geq}t\sim \pi}[R_t|s_t,a_t].
\end{align}

If the target policy is deterministic, it can be described as a function $\mu :\mathcal{S}\rightarrow \mathcal{A}$. The Bellman equation in Q-learning \cite{watkins92} can be expressed as:
\begin{align}
Q^\mu(s_t,a_t) = \mathbb{E}_{r_t,s_{t+1}\sim E}[r_t(s_t,a_t)+\gamma Q^\mu(s_{t+1},\mu(s_{t+1}))].
\end{align}
Parameterized the function approximators by $\theta^Q$, the parameters/weights can be optimized by minimizing the loss:
\begin{align}
L(\theta ^Q) = \mathbb{E}\big[(y_t - Q(s_t,a_t|\theta^Q))^2 \big],
\label{q-learning}
\end{align}
where $y_t = r(s_t,a_t)+\gamma Q(s_{t+1},\mu (s_{t+1})|\theta ^Q)$.

\subsection{Deep Deterministic Policy Gradient Algorithm}
\begin{figure}[htbp]
	\centerline{\includegraphics[width=0.47\textwidth]{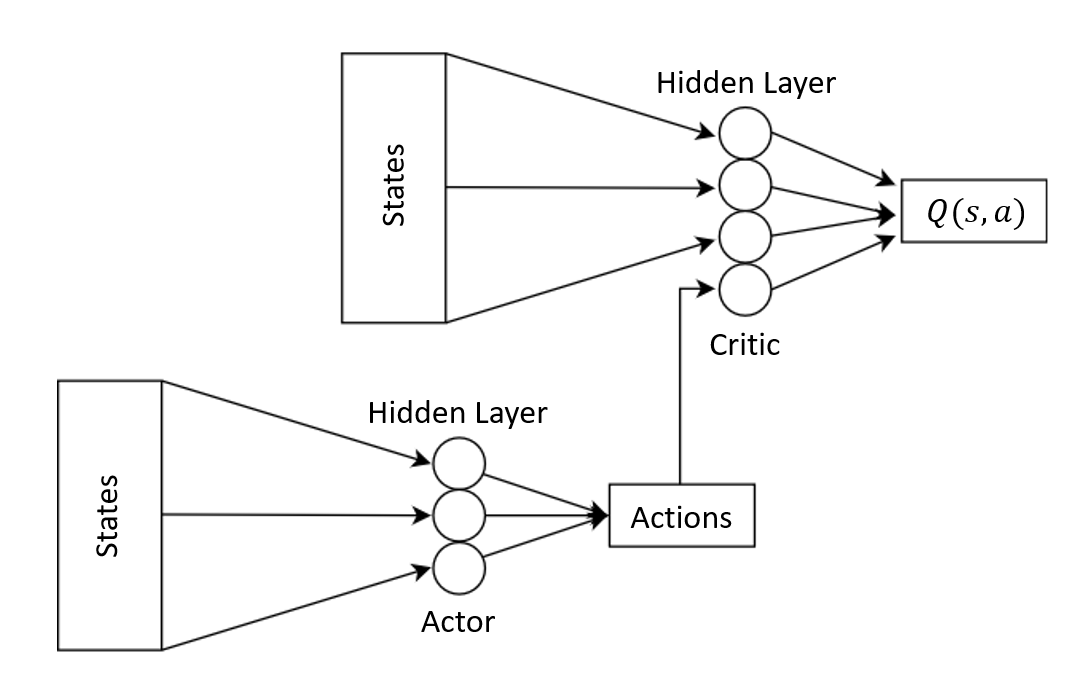}}
	\caption{Deep deterministic policy gradient network.}
	\label{fig:ddpg}
\end{figure}

Applying Q-learning to continuous action space is problematic since the greedy policy requires global optimization during policy improvement. Deterministic policy gradient (DPG) is more computationally tractable for problems over continuous action space \cite{silver2014}. The DPG keeps a parameterized actor function $\mu (s|\theta ^\mu)$. The critic $Q(s,a)$ is learned based on Bellman equation as in Q-learning. Fig.~\ref{fig:ddpg} shows the structure of the deterministic actor critic network. The actor is updated via gradient descent to maximize the expected return from the start distribution $J$:
\begin{align}
\nabla _{\theta ^\mu}\approx \mathbb{E}[\nabla_{\theta ^\mu} Q(s,a|\theta^Q)|s=s_t,a=\mu(s_t|\theta ^\mu)]. 
\end{align}

In this paper, a similar approach is adapted from \cite{timothy2016}, which uses deep neural networks as function approximators for DPG. This approach is referred as deep deterministic policy gradient (DDPG), which outperforms other continuous action algorithms \cite{smruti2017}.
Several techniques are applied to improve the performance of DDPG:
\begin{itemize}
	\item \textbf{Replay buffer}: the replay buffer is also called experience relay, which is a finite sized memory buffer $\mathcal{R}$. The state reward tuple ($s_t,a_t,r_t,s_{t+1}$) is stored in $\mathcal{R}$. The oldest samples are discarded when the buffer is full. A minibatch is uniformly sampled from the buffer to update the actor and critic. The replay buffer allows the algorithm to learn across uncorrelated transitions and also more efficient usage of hardware.
	\item \textbf{Target network}: implementing Q-learning (\ref{q-learning}) with neural networks is prone to divergence since the network $Q(s,a|\theta ^Q)$ is also being updated when calculating the target value $y_t$. The solution used is similar to the target network used in \cite{mnih2013} with some modifications. Instead of directly copying weights of the learned networks as in \cite{mnih2013}, the actor and critic target networks for actor and critic in this paper ($\mu ^\prime (s|\theta ^{\mu^\prime})$, $Q^\prime(s,a|\theta ^{Q^\prime})$) are allowed to slowly track the learned networks: $\theta ^\prime \leftarrow \tau \theta +(1-\tau)\theta ^\prime$ with $\tau \ll 1$. 
	\item \textbf{Batch normalization}: when dealing with data of physical systems, different components may have different units and their ranges may vary across different environments, which makes it difficult for the neural network to learn effectively.  Batch normalization is a technique introduced in \cite{ioffe2015} to deal with this issue. It normalizes all dimensions of samples within a minibatch to have unit mean and variance. In deep neural networks, batch normalization can also prevent covariance shift, therefore enabling usage of larger learning rate.
	\item \textbf{Ornstein-Uhlenbeck exploration}: exploration is a major challenge for learning in continuous action space. The exploration policy $\mu^\prime$ is constructed by adding noise sampled from a noise process $\mathcal{M}$:
	\begin{align}
	\mu ^\prime(s_t)=\mu(s_t|\theta_t^\theta)+\mathcal{M}.
	\end{align}
	The noise $\mathcal{M}$ is generated from Ornstein-Uhlenbeck process \cite{uhlenbeck1930}, which is temporally correlated for exploration efficiency in physical control problem with inertia.
\end{itemize}

\subsection{Implementation of DDPG for Smart Inverter Coordination}
\label{sec:implementation}

\begin{figure}[htbp]
	\centerline{\includegraphics[width=0.45\textwidth]{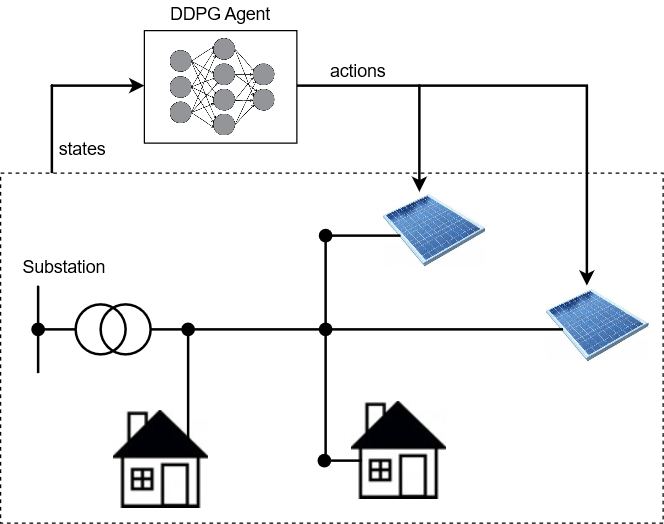}}
	\caption{Schematic overview of using a trained DDPG agent for SI coordination (test/deployment phase). Sensors provide real-time measurements/states of the power grid, which are fed into a well-trained DDPG agent. The DDPG agent makes decisions of actions SIs should take. This process requires only one feed-forward step of the trained neural network and therefore is very fast. The actions suggested by the DDPG agent are sent back to SIs to execute.}
	\label{fig:grid}
\end{figure}

The goal of a well-trained DDPG agent for SI coordination is to provide fast yet effective actions for ensuring normal voltage performance and minimization of PV production curtailment. The actions are determined based on real-time measurements (states) from the power grid as shown in Fig.~\ref{fig:grid}. 

Before applying the DDPG agent for gird control, it needs to be properly trained. The definition of episode, state, action and reward is provided below:

\textit{1): Episode}

An episode is any operation scenario collected from a real-time measurement system, e.g. phasor measurement units (PMUs), under random PV production fluctuations and load changes. In this work, only steady state is studied without considering transients.

\textit{2): State Space}

The state $s$ is defined as a vector containing power system information, including voltage magnitudes of each bus, real and reactive power generation/consumption of PVs and loads. The state space $\mathcal{S}$ is the space $s$ belongs to.

\textit{3): Action Space}

In this work, we consider SI reactive power outputs as actions. Allowing PV real power curtailment, each SI can adjust its reactive power output from $-S$ to $S$ (Fig.~\ref{fig:SI}). Considering normalized output, each SI can alter its reactive power continuously within [-1,1] p.u. \footnote{All system voltage, real power and reactive power in this paper are presented in per unit (p.u.) values, which are their actual values divided by  nominal base values.}. The action space $\mathcal{A}$ is spanned by action combinations of all SIs.

\textit{4): Reward}

When applying RL to control, the reward scheme needs to be carefully designed to achieve proper system performance. In this paper, the objectives are mitigation of voltage violations and minimization of PV generation curtailment. The reward scheme is also  composed of two parts to meet these two goals: the large penalty for violating voltage limits and the negative reward proportional to total reactive power dispatched by SIs.

\begin{figure}[htbp]
	\centerline{\includegraphics[width=0.5\textwidth]{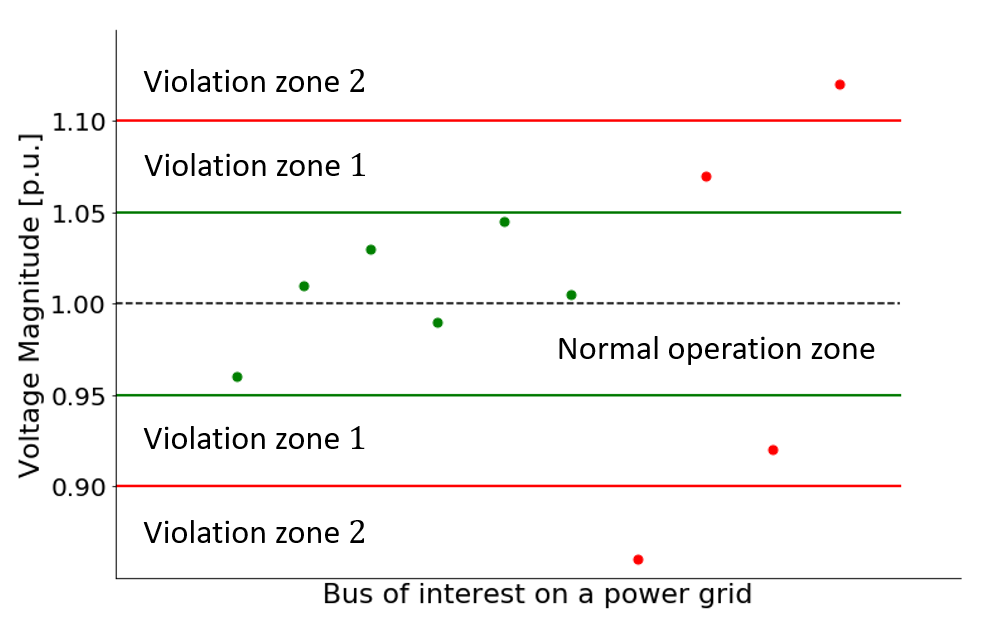}}
	\caption{Voltage profile zone definition.}
	\label{fig:reward_scheme}
\end{figure}

The first part of the reward is assigned according to voltage profiles.
Several voltage operation zones are defined for differentiating system voltage profiles, including normal zone (0.95 - 1.05) p.u., violation zone 1 (0.9 - 0.95 or 1.05 - 1.1 ) p.u., and violation zone 2 ($<$ 0.9 or $>$ 1.1) p.u., as presented in Fig.~\ref{fig:reward_scheme}. Those zones are defined according to the grid operation limits \cite{ansi}. For one episode, let's assume $|V_k|$ is the voltage magnitude at bus $k$. The reward associated with $|V_k|$ for bus $k$ in $j^{th}$ iteration/exploration step can be calculated as, 

\begin{align}
R_V(j,k) = \begin{cases}
0, &  \text{if }  {  } |V_k| \in \text{normal zone} \\
-400, & \text{if } {  }  |V_k| \in \text{violation zone 1}  \\
-600, & \text{if } {  } |V_k| \in \text{violation zone 2}.
\end{cases}
\label{eq:reward_volt}
\end{align}

The second part of the reward is assigned based on the reactive power utilization. Besides ensuring proper voltage profiles, the other objective is to minimize the PV production curtailment, which is assured by minimization of reactive power utilization. The reward for reactive power utilization is defined as follows,
\begin{align}
R_Q(j) = \sum_{i=1}^{M}C\times(1-q_i),
\end{align}
where $q_i = |Q_i|/S_i$ is the reactive power utilization ratio of $i^{th}$ SI (i.e. the absolute value of action for the SI); $M$ is the total number of SI; $C$ is a constant chosen to scale the reward, which is set to be 200 in this work. The value of $C$ needs to be tuned to fit different power system configurations for desirable performance.
The total reward for $j^{th}$ iteration/exploration step of the episode is:
\begin{align}
R(j) = \sum_{k=0}^{N}R_V(j,k) + R_Q(j)
\label{eq:reward_iter},
\end{align}
where $\{0,...,N\}$ are the indexes for all buses as stated in Section \ref{sec:prel1}.

\subsection{Training of DDPG}
\label{sec:training}

\begin{figure}[htbp]
	\centerline{\includegraphics[width=0.48\textwidth]{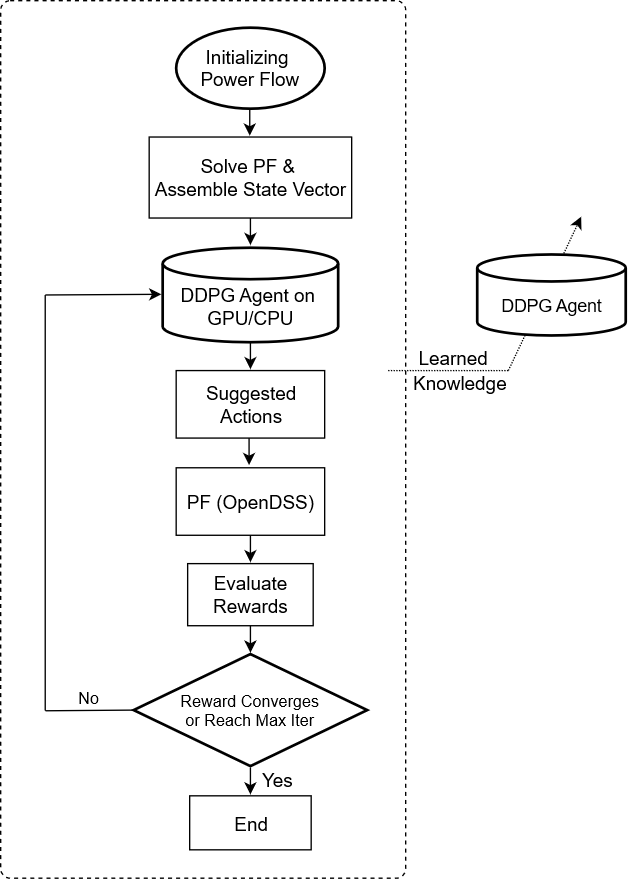}}
	\caption{Flowchart of training a DDPG agent.}
	\label{fig:flowchart}
\end{figure}

With key concepts defined above, training of the DDPG can be done following the procedures displayed in Fig.~\ref{fig:flowchart}. The training consists of the following key steps:

\textbf{Step 1}: at the beginning of one episode (one operation scenario of the power network), the power flow (PF) will be solved to get the system information and assemble the state vector. The PF is performed by the AC power flow solver OpenDSS \cite{opendss}, which takes in load consumption and PV generation information, solves the corresponding PF equations (\ref{pf_eq1},\ref{pf_eq2}), and gets voltage at each bus.

\textbf{Step 2}: the state vector containing the system information (bus voltage, real and reactive power consumption/generation of SIs and loads) is fed into the DDPG agent. The agent generates suggested actions, which are reactive power outputs of SIs.

\textbf{Step 3}: the environment (e.g. the AC power flow solver) takes the suggested actions and produce the resultant state by solving another PF. The corresponding reward for that state is evaluated. If the termination is reached according to the termination criteria defined below, the training for this episode is terminated and the trained DDPG is stored for later use. 

\textbf{Step 4}: if the termination criteria is not met, return to \textbf{Step 2}.

The training for one episode terminates if :1) the reward for the exploration/iteration step converges, meaning the reward difference is less than 5 for five consecutive steps (the convergence is determined after 200 iteration steps for each episode); 2) the maximum number of iterations (1000) is reached.

\section{Case Study and Discussion}
\label{sec:casestudy}
\subsection{Case Study}

The performance of the proposed DDPG agent is tested on a modified IEEE 37 node test feeder. The properties of the test feeder is summarized in Table.~\ref{table:ieee37}. Five 1.2 MW PVs (totaling 6 MW) are added to the feeder. The AC rating of each SI is 1 MVA, assuming 20\% oversizing of DC solar panel \cite{van2003}. Three different cases are studied: 1) \textbf{Baseline}, the SI operates at unity power factor without reactive power generation; 2) \textbf{Volt-Var} (benchmark), the SI operates according to local information autonomously as depicted in Fig.~\ref{fig:voltvar}; 3) \textbf{DDPG} (proposed), SIs are coordinated following the decisions made by the well-trained DDPG agent as described in Section~\ref{sec:DRL}.

The training of the DDPG agent is performed following the procedures shown in Fig.~\ref{fig:flowchart}. In the training stage, PV generation and load consumption combinations are randomly generated to represent different categories of grid operation conditions :1) high load and no PV production for evening periods, when the grid is prone to under-voltages; 2) low load and high PV production for rare middle-day intervals, during which over-voltages are more likely to occur; 3) moderate load and PV production for normal day-time scenarios.

\begin{figure}[htbp]
	\centerline{\includegraphics[width=0.48\textwidth]{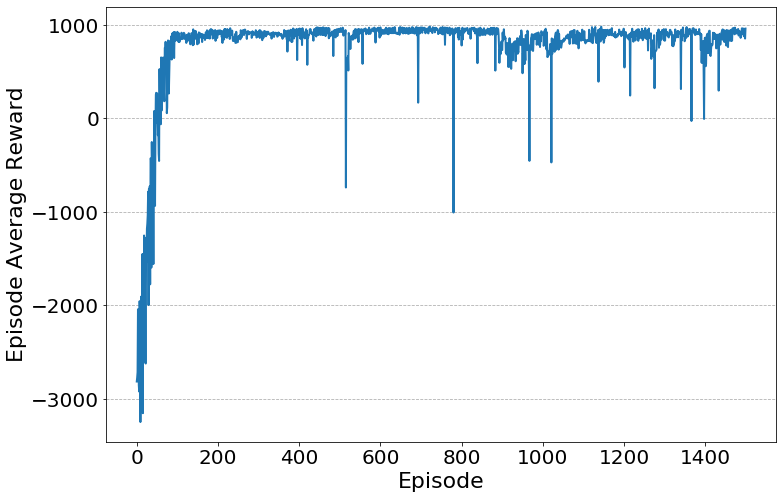}}
	\caption{Episode average reward during training process. The training is performed for 1500 episodes.}
	\label{fig:reward}
\end{figure}

Training is performed for 1500 episodes with a total iterations of approximately 500 k, i.e. each episode terminates after approximately 330 iterations on average. The average episode reward is plotted in Fig.~\ref{fig:reward}. With the test case setup, the maximum reward the DDPG agent can get is 1000. The reward starts at a low value, given that the DDPG agent has zero prior knowledge on how to perform grid voltage regulation. The DDPG agent is very efficient of learning from past experiences and gets over 800 of reward value after just 100 episodes. As learning progresses, the reward slowly increases to over 900. There are sudden dips of the reward during the training, it's likely caused by two factors: 1) the DDPG agent experiences a complicated grid condition not seen before; 2) the DDPG takes unusual actions for exploration. During later phase of the training, the sudden dips become smaller, indicating converging of the model.     

\begin{figure}[htbp]
	\centerline{\includegraphics[width=0.48\textwidth]{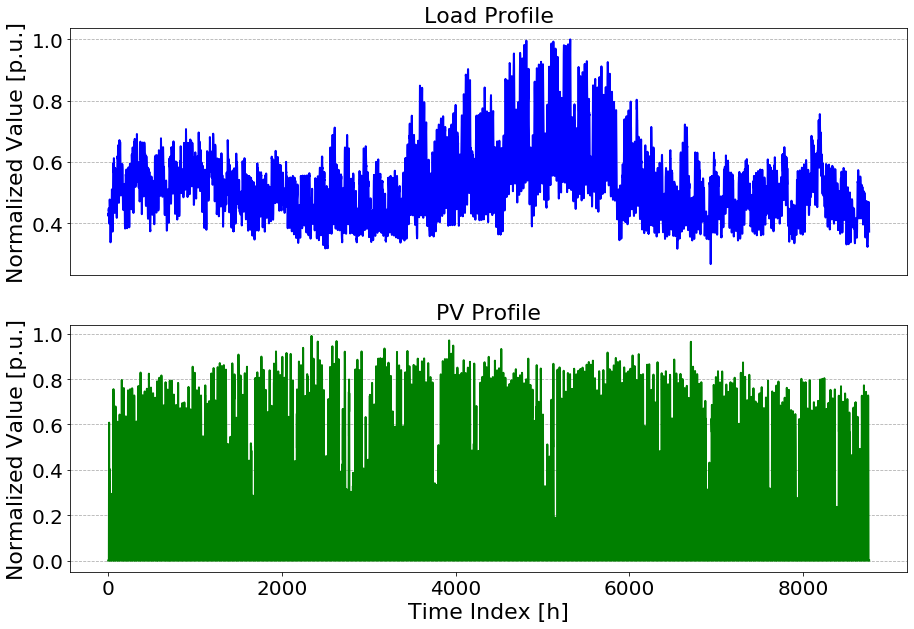}}
	\caption{Normalized load (top) and PV (bottom) profiles used for tests. The load profile is from OpenDSS \cite{opendss} dataset. The PV generation profile is from public solar datesets maintained by NREL \cite{solardata}}.
	\label{fig:ld_solar}
\end{figure}

After training for 1500 episodes, the DDPG agent is used to perform grid voltage control. one year comprehensive tests are done with 1 hour resolution (8760 different scenarios) to evaluate the agent's performance. To make the task harder, online reward feedbacks of the grid after taking the suggested actions are not used to retrain the DDPG agent during the test (as shown in Fig.~\ref{fig:grid}). Therefore, the DDPG agent makes decision solely based on past experiences learned during the training phase. The PV generation and load consumption profiles used for the test are plotted in Fig.~\ref{fig:ld_solar}. 

\begin{table}[]
	\caption{Properties of modified IEEE 37 node test feeder.}
	\centering
	\begin{tabular}{l|c}
		\hline
		\# of Nodes           & 37   \\ \hline
		Peak Load (MVA)       & 2.74 \\ \hline
		\# of Loads           & 25   \\ \hline
		\# of PVs             & 5    \\ \hline
		DC Rating of PVs (MW)  & 6    \\ \hline
		AC Rating of SIs (MVA) & 5    \\ \hline
	\end{tabular}
	\label{table:ieee37}
\end{table}

\subsection{Results and Discussion}

\begin{table*}[ht]
	\caption{Summary of test results. The value shown is the accumulated quantity of each parameter for 1 year.}
	\begin{center}
		\begin{tabular}{l|c|c|c|c}
			\hline
			& \# of Under-voltages & \multicolumn{1}{l|}{\# of Over-voltages} & \multicolumn{1}{l|}{PV Curtailment (kWh)} & \multicolumn{1}{l}{System Losses (kWh)} \\ \hline
			Baseline             & 630                  & 38,201                                   & 0                                         & 394,131                                 \\ \hline
			Volt-Var (benchmark) & 0                    & 0                                        & 14,610                                    & 440,187                                 \\ \hline
			DDPG (proposed)      & 0                    & 0                                        & 3,330                                     & 396,015                                 \\ \hline
		\end{tabular}
	\end{center}
	\label{table:results}
\end{table*}

\begin{figure}[htbp]
	\centerline{\includegraphics[width=0.49\textwidth]{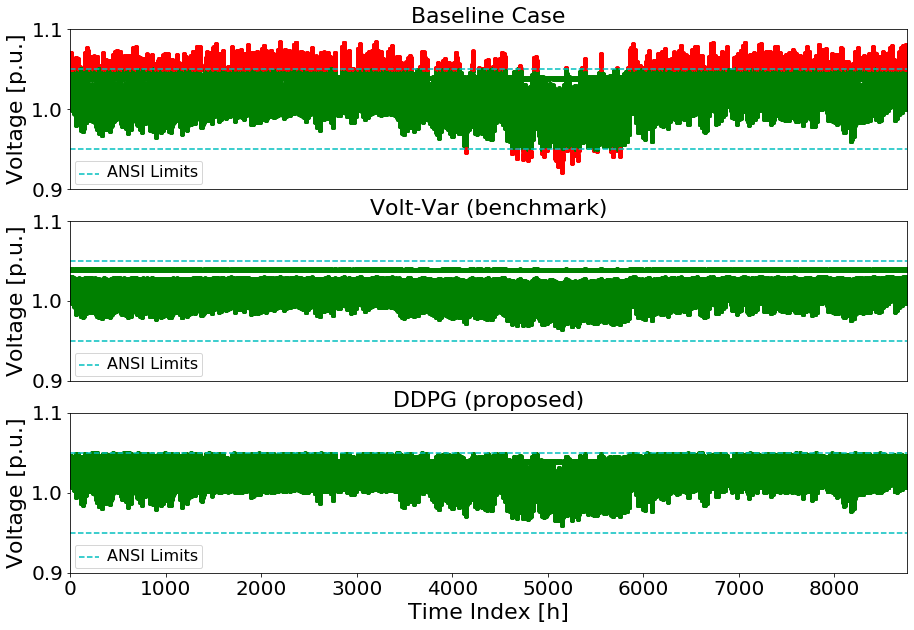}}
	\caption{Voltage profiles of 1 year test for three cases. Each dot represents the voltage of one node at one time step. Green dots indicate normal nodal voltages within [0.95,1.05] p.u. ANSI limits \cite{ansi}. Red dots mark nodal voltages out of ANSI limits, which are represented by cyan dash lines.}
	\label{fig:voltage}
\end{figure}

Fig.~\ref{fig:voltage} displays voltage profiles of three cases. Numerous voltage violations can be observed for baseline case, most of which are over-voltages and some are under-voltage violations. Over-voltages typically happen during sunny middle-day when excessive PV generation causes reverse power flow, leading to voltage rise on the distribution grid. Without reactive power support from SIs, the grid is prone to over-voltage issues. Under-voltages are more likely in peak evening hours, when demand is high and PV production is zero. During these intervals, large voltage drop along the feeder makes under-voltage violations more likely. Zero voltage violation is present for both Volt-Var and DDPG cases. Interestingly, a number of green dots are very close to the 1.05 p.u. upper limit without crossing it in DDPG case. The maximum voltage observed in DDPG case is 1.0494 p.u.. This shows the DDPG agent learned a delicate strategy to utilize minimal amount of reactive power to keep the voltage just below the limit. This demonstrates the voltage constraints posed by penalty as described in Section \ref{sec:implementation} are effective.

\begin{figure}[htbp]
	\centerline{\includegraphics[width=0.49\textwidth]{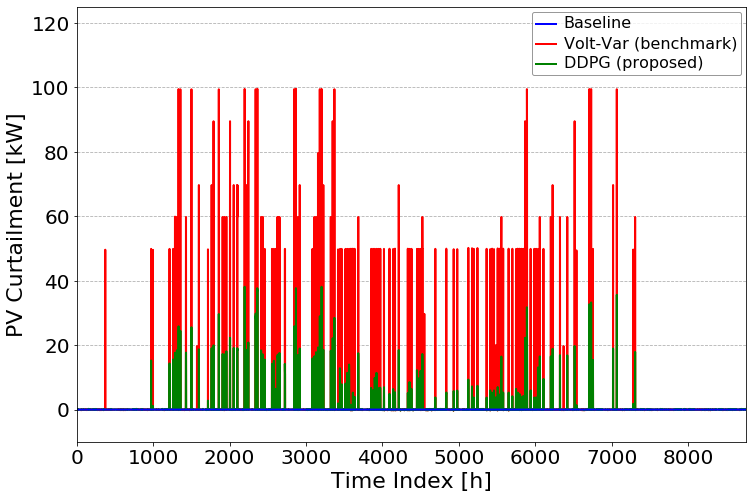}}
	\caption{PV generation curtailment for three cases. The pv curtailment here is defined as the PV real power generation deficit between baseline case (no reactive power utilization) and other two cases (with reactive power generation). Therefore, the PV curtailment for baseline case is always zero. In this way, the curtailment resulted from reactive power utilization of the SI is quantified, providing a direct comparison of Volt-Var and DDPG on effective usage of SI reactive power.}
	\label{fig:curtailment}
\end{figure}

The curtailment of PV production due to reactive power utilization is displayed in Fig.~\ref{fig:curtailment}. Since reactive power usage is prohibited in baseline case, the corresponding curtailment is always zero. Comparing to baseline case, the real power of the SI needs to be curtailed to make room for reactive power generation (Fig.~\ref{fig:SI}) in both Volt-Var and DDPG cases. However, since the DDPG can coordinates different SIs to utilize reactive power more efficiently, much less curtailment is incurred comparing to Volt-Var case. The total energy curtailed for the DDPG case is only 23$\%$ of Volt-Var case (Table.~\ref{table:results}), achieving 77$\%$ reduction in curtailment.

Improper reactive power injection/absorption may lead to increased system losses. The system losses of all three cases are summarized in Table.~\ref{table:results}. The system loss of baseline case is lowest. Volt-Var increases the system loss by 46,056 kWh comparing to  baseline case while the DDPG only causes a marginal 1,884 kWh increment, which is only 4.1$\%$ of Volt-Var case in term of additional losses resulted from reactive power generation.

\section{Conclusion}
\label{sec:conc}

In this paper, a DDPG based algorithm is proposed to coordinate multiple SIs for distribution grid voltage regulation. The DDPG can handle continuous control of SIs and provide fast coordination decisions. Comprehensive tests with thousands of realistic scenarios are conducted on the IEEE 37 node feeder to evaluate performance of a well-trained DDPG agent. The DDPG agent is compared against baseline case and  Volt-Var benchmark case. The results demonstrate that even without online reward feedbacks from the grid, a well-trained DDPG agent can use knowledge learned from the training phase to make robust decisions, which can effectively mitigate voltage violations, significantly reduce PV production curtailment, and substantially decrease additional system losses incurred by reactive power utilization.

%

\begin{thebibliography}{10}
\providecommand{\url}[1]{#1}
\csname url@samestyle\endcsname
\providecommand{\newblock}{\relax}
\providecommand{\bibinfo}[2]{#2}
\providecommand{\BIBentrySTDinterwordspacing}{\spaceskip=0pt\relax}
\providecommand{\BIBentryALTinterwordstretchfactor}{4}
\providecommand{\BIBentryALTinterwordspacing}{\spaceskip=\fontdimen2\font plus
\BIBentryALTinterwordstretchfactor\fontdimen3\font minus
  \fontdimen4\font\relax}
\providecommand{\BIBforeignlanguage}[2]{{%
\expandafter\ifx\csname l@#1\endcsname\relax
\typeout{** WARNING: IEEEtran.bst: No hyphenation pattern has been}%
\typeout{** loaded for the language `#1'. Using the pattern for}%
\typeout{** the default language instead.}%
\else
\language=\csname l@#1\endcsname
\fi
#2}}
\providecommand{\BIBdecl}{\relax}
\BIBdecl

\bibitem{walling2008}
R.~Walling, R.~Saint, R.~C. Dugan, J.~Burke, and L.~A. Kojovic, ``Summary of
  distributed resources impact on power delivery systems,'' \emph{IEEE
  Transactions on power delivery}, vol.~23, no.~3, pp. 1636--1644, 2008.

\bibitem{li2018optimal}
C.~Li, V.~R. Disfani, Z.~K. Pecenak, S.~Mohajeryami, and J.~Kleissl, ``Optimal
  oltc voltage control scheme to enable high solar penetrations,''
  \emph{Electric Power Systems Research}, vol. 160, pp. 318--326, 2018.

\bibitem{ieee1547_2}
\BIBentryALTinterwordspacing
``{IEEE} standard for interconnection and interoperability of distributed
  energy resources with associated electric power systems interfaces.''
  [Online]. Available: \url{https://doi.org/10.1109/ieeestd.2018.8332112}
\BIBentrySTDinterwordspacing

\bibitem{ca21}
{California Public Utilities Commission}, ``Rule 21 interconnection,'' 2018.

\bibitem{rule14h}
{Hawaiian Electric}, {Maui Electric}, and {Hawaii Electric Light}, ``Hawaiian
  electric rules,'' 2018.

\bibitem{emiliano14}
E.~Dall’Anese, S.~V. Dhople, and G.~B. Giannakis, ``Optimal dispatch of
  photovoltaic inverters in residential distribution systems,'' \emph{IEEE
  Transactions on Sustainable Energy}, vol.~5, no.~2, pp. 487--497, 2014.

\bibitem{guggilam2016}
S.~S. Guggilam, E.~Dall’Anese, Y.~C. Chen, S.~V. Dhople, and G.~B. Giannakis,
  ``Scalable optimization methods for distribution networks with high pv
  integration,'' \emph{IEEE Transactions on Smart Grid}, vol.~7, no.~4, pp.
  2061--2070, 2016.

\bibitem{li2019}
C.~Li, V.~Disfani, H.~V. Haghi, and J.~Kleissl, ``Optimal voltage regulation of
  unbalanced distribution networks with coordination of oltc and pv
  generation,'' in \emph{2019 IEEE Power and Energy Society general meeting},
  2019.

\bibitem{alphago17}
D.~Silver, J.~Schrittwieser, K.~Simonyan, I.~Antonoglou, A.~Huang, A.~Guez,
  T.~Hubert, L.~Baker, M.~Lai, A.~Bolton \emph{et~al.}, ``Mastering the game of
  go without human knowledge,'' \emph{Nature}, vol. 550, no. 7676, p. 354,
  2017.

\bibitem{arcade13}
M.~G. Bellemare, Y.~Naddaf, J.~Veness, and M.~Bowling, ``The arcade learning
  environment: An evaluation platform for general agents,'' \emph{Journal of
  Artificial Intelligence Research}, vol.~47, pp. 253--279, 2013.

\bibitem{robotics15}
J.~Kober, J.~A. Bagnell, and J.~Peters, ``Reinforcement learning in robotics: A
  survey,'' \emph{The International Journal of Robotics Research}, vol.~32,
  no.~11, pp. 1238--1274, 2013.

\bibitem{diao19}
R.~Diao, Z.~Wang, D.~Shi, Q.~Chang, J.~Duan, and X.~Zhang, ``Autonomous voltage
  control for grid operation using deep reinforcement learning,'' \emph{arXiv
  preprint arXiv:1904.10597}, 2019.

\bibitem{yang19}
Q.~Yang, G.~Wang, A.~Sadeghi, G.~B. Giannakis, and J.~Sun, ``Real-time voltage
  control using deep reinforcement learning,'' \emph{arXiv preprint
  arXiv:1904.09374}, 2019.

\bibitem{xu12}
Y.~Xu, W.~Zhang, W.~Liu, and F.~Ferrese, ``Multiagent-based reinforcement
  learning for optimal reactive power dispatch,'' \emph{IEEE Transactions on
  Systems, Man, and Cybernetics, Part C (Applications and Reviews)}, vol.~42,
  no.~6, pp. 1742--1751, 2012.

\bibitem{xu18}
H.~Xu, A.~D. Dom{\'\i}nguez-Garc{\'\i}a, and P.~W. Sauer, ``Optimal tap setting
  of voltage regulation transformers using batch reinforcement learning,''
  \emph{arXiv preprint arXiv:1807.10997}, 2018.

\bibitem{wei2019}
W.~Wang, N.~Yu, J.~Shi, and Y.~Gao, ``Volt-var control in power distribution
  systems with deep reinforcement learning,'' \emph{IEEE SmartGridComm}, 2019.

\bibitem{john2004}
J.~G. Vlachogiannis and N.~D. Hatziargyriou, ``Reinforcement learning for
  reactive power control,'' \emph{IEEE transactions on power systems}, vol.~19,
  no.~3, pp. 1317--1325, 2004.

\bibitem{timothy2016}
T.~P. Lillicrap, J.~J. Hunt, A.~Pritzel, N.~Heess, T.~Erez, Y.~Tassa,
  D.~Silver, and D.~Wierstra, ``Continuous control with deep reinforcement
  learning,'' \emph{arXiv preprint arXiv:1509.02971}, 2015.

\bibitem{watkins92}
C.~J. Watkins and P.~Dayan, ``Q-learning,'' \emph{Machine learning}, vol.~8,
  no. 3-4, pp. 279--292, 1992.

\bibitem{silver2014}
D.~Silver, G.~Lever, N.~Heess, T.~Degris, D.~Wierstra, and M.~Riedmiller,
  ``Deterministic policy gradient algorithms,'' 2014.

\bibitem{smruti2017}
S.~Amarjyoti, ``Deep reinforcement learning for robotic manipulation-the state
  of the art,'' \emph{arXiv preprint arXiv:1701.08878}, 2017.

\bibitem{mnih2013}
V.~Mnih, K.~Kavukcuoglu, D.~Silver, A.~Graves, I.~Antonoglou, D.~Wierstra, and
  M.~Riedmiller, ``Playing atari with deep reinforcement learning,''
  \emph{arXiv preprint arXiv:1312.5602}, 2013.

\bibitem{ioffe2015}
S.~Ioffe and C.~Szegedy, ``Batch normalization: Accelerating deep network
  training by reducing internal covariate shift,'' \emph{arXiv preprint
  arXiv:1502.03167}, 2015.

\bibitem{uhlenbeck1930}
G.~E. Uhlenbeck and L.~S. Ornstein, ``On the theory of the brownian motion,''
  \emph{Physical review}, vol.~36, no.~5, p. 823, 1930.

\bibitem{ansi}
A.~Std, ``C84. 1-2011,'' \emph{American National Standard for Electric Power
  Systems and Equipment-Voltage Ratings (60 Hertz)}, 2011.

\bibitem{opendss}
R.~C. Dugan, ``The open distribution system simulator (opendss),'' \emph{EPRI
  OpenDSS Manual}, 2012.

\bibitem{van2003}
N.~Van Der~Borg and A.~Burgers, ``Inverter undersizing in pv systems,'' in
  \emph{3rd World Conference on Photovoltaic Energy Conversion, 2003},
  vol.~2.\hskip 1em plus 0.5em minus 0.4em\relax IEEE, 2003, pp. 2066--2069.

\bibitem{solardata}
{GE Energy}, ``Western wind and solar integration study,'' No.
  NREL/SR-550-47434. National Renewable Energy Lab.(NREL), Golden, CO (United
  States), Tech. Rep., 2010.

\end{thebibliography}


\end{document}